\documentstyle[pra,preprint,aps]{revtex}
\begin{document}

\title{ Perfect Teleportation and Superdense Coding With W-States}

\author{Pankaj Agrawal and Arun Pati }
\address{Institute of Physics, Bhubaneswar-751005, Orissa, India.}

\newcommand{\sqrttwo}{{1 \over \sqrt{2} }}
\newcommand{\sqrtthr}{{1 \over \sqrt{3} }}
\newcommand{\half}{{1 \over 2 }} 
\newcommand{\quart}{{1 \over 4 }} 
\newcommand{\norm}{{1 \over \sqrt{2 + 2 n  } }}
\newcommand{\Non}{{1 \over \sqrt{1 + |n|^2  } }}
\newcommand{\Nop}{{1 \over \sqrt{1 + |p|^2  } }}
\newcommand{\Nolp}{{1 \over \sqrt{1 + |\ell^{\prime}|^ 2} }}
\newcommand{\Nom}{{1 \over \sqrt{1 + |m|^2    } }}
\newcommand{\Nopp}{{1 \over \sqrt{1 + |p^{\prime}|^2    } }}

\newcommand{\be}{\begin{equation}}
\newcommand{\ee}{\end{equation}}
\newcommand{\bea}{\begin{eqnarray}}
\newcommand{\eea}{\end{eqnarray}}

\newcommand{\rag}{\rangle}
\newcommand{\lag}{\langle}
\def\ie{\hbox{\it i.e.}{}}
\def\eg{\hbox{\it e.g.}{}}

\maketitle
\def\ra{\rangle}
\def\la{\langle}
\def\ver{\arrowvert}

\maketitle

\begin{abstract}
     True tripartite entanglement of the state of a system of 
     three qubits can 
     be classified on the basis of stochastic local operations 
     and classical communications (SLOCC). Such states can be
     classified in two categories: GHZ states and W-states.
     It is known that GHZ states can be used for teleportation
     and superdense coding, but the prototype W-state cannot be. However, 
     we show that there is a class of W-states that can 
     be used for perfect teleportation and superdense coding.
\end{abstract}

\vskip 1cm

PACS           NO:    03.67.-a, 03.65.Bz\\

\vskip 12cm

$\overline{{\rm Emails: 
agrawal@iopb.res.in,~ akpati@iopb.res.in }}$

\vfill



\vspace{.1in}

\section{Introduction}

Quantum teleportation is a prime example of 
quantum information processing task where an unknown state can be 
perfectly transported from one place to another using previously shared
entanglement and classical communication between the sender and the receiver.
This is a remarkable application of entangled states which has many 
ramifications in quantum information technology. The original protocol 
of Bennett {\em et~al} \cite{bbcjpw} involves teleportation 
of an unknown qubit using an EPR pair and by sending two bits 
of classical information from Alice to Bob. They have 
also generalized the protocol for an unknown qudit 
using maximally entangled state in $d\times d$ dimensional Hilbert space
and by sending $2\log_{2} d$ bits
of classical information. Then, quantum teleportation was generalized to the 
case where sender and receiver do not have perfect EPR pair but noisy channel
 \cite{bbps}.
Quantum teleportation is also possible for infinite dimensional Hilbert 
spaces, so called continuous variable quantum teleportation \cite{lv,bk}. 
If we do not have maximally entangled state then one cannot teleport a qubit
with unit fidelity and unit probability. However, it is possible to have unit 
fidelity teleportation but with a probability less than unit--called 
probabilistic quantum teleportation \cite{ap,pa}. 
Using non-maximally entangled 
basis as a measurement basis this was shown to be possible.
Subsequently, this
probabilistic scheme has been generalized to teleport $N$ qubits \cite{gordon}.
In the past several years quantum teleportation has been also experimentally 
verified by various groups \cite{db,dbe,af}.

Quantum entanglement lies at the heart of quantum teleportation and,
in fact, at other host of information processing tasks. Since this refers
to the property of multiparticle (two or more) quantum states, a natural 
question that comes to mind is that in addition to the existing 
two-particle entangled states whether one can exploit other multiparticle 
entangled states for quantum teleportation. That one can teleport an unknown 
qubit using three particle GHZ state \cite{kb} and four particle GHZ state 
\cite{akp} was shown 
to be possible. Dealing with multiparticle entangled states is not always
easy. In case of three qubits, the class of entangled states have been well 
studied. There are two types of inequivalent entangled states for three qubits,
one is GHZ type and other is W-type. It has been shown by 
Duer {\em et al} \cite{dvc} that W-states cannot be converted to GHZ 
states under stochastic local operation and
classical communication (SLOCC). There is an interesting class 
of three qubit entangled states called zero sum amplitude (ZSA) states
which are useful for creating universal entangled states \cite{pati}. 
These are not equivalent to GHZ states. It may be mentioned that
the GHZ states are not robust against loss of particles (i.e., if we 
trace out one particle then there is no entanglement between remaining 
two) while W-states are. If we trace out any one particle 
from W-state, then there is some genuine entanglement between the 
remaining two.  Therefore, in contrast to a GHZ state, a W-state can
still be used as a resource even after the loss of a particle.
Recently, the W-state has been used for quantum key distribution \cite{joo}
and in illustrating violation of local realism in an interesting way that is
not seen in GHZ state \cite{cabe}.
 Considering the importance of W-states, there have been 
various proposals to prepare W-states in the literature \cite{guo,gor,ba}.

Entangled states not only enhance our ability to do variety of quantum 
information processing tasks but they also enhance classical information 
capacity. For example, in superdense coding if Alice and Bob share a
EPR pair, then by sending a qubit Alice can send two classical bits  
\cite{bw}. More generally, if Alice and Bob share a maximally entangled 
state in $d \times d$ dimensional Hilbert space, then by sending a qudit 
Alice can communicate $2\log_{2} d$ 
classical bits. Thus, a maximally entangled state doubles the classical 
information capacity of a channel. Similar to quantum teleportation, 
in recent years, people have exploited 
other class of entangled states for dense coding. One can also use 
non-maximally entangled state for superdense coding either in a deterministic
fashion \cite{haus} or in a probabilistic manner \cite{hao,ppa}. 
Recently, many people have explored various kind of quantum channels 
for determinstic and unambiguous superdense coding \cite{wu,fan,feng}. 

The purpose of this paper is to show that there exist a class of W-state 
which can be used for perfect quantum teleportation and superdense 
coding. Earlier, quantum teleportation protocol using W-state have 
been studied \cite{gor1}. But it was shown that one needs to do 
non-local operation to recover the unknown state. In another piece of
work it has been shown that if one uses W-state then the teleportation 
protocol works with non-unit fidelity, i.e., it is not perfect \cite{jpok}.
Also, it is not possible to recover the unknown state using W-state as a
channel. However, in the present work we show that by performing von 
Neumann projection on the three-particles and by sending two bits of 
classical information one can teleport an unknown qubit perfectly using 
a class of W-states. Furthermore, we show how one can perform superdense 
coding with the same class of W-states. Interestingly, the quantum resource 
used in teleportation and dense coding protocols presented here 
with W-states is again one ebit of shared entanglement between Alice and Bob.

The organization of this paper is as follows. In section II, we discuss a class
of W-states that are useful for quantum teleportation and dense coding. 
In section III, we discuss the actual teleportation protocol. For the sake of
completeness we also discuss the teleportation protocol using GHZ state. 
In section IV, we discuss superdense coding protocol using W-state. Then, the 
conclusion follows in section V.

\vspace{0.05in}

\section{ A Class of W-states}

In this section we briefly discuss a class of W-states that are useful for 
perfect quantum teleportation and superdense coding.
It may be recalled that although prototype GHZ-state such as 
\be
|GHZ \rag_{123} = \sqrttwo \,(|000\rag_{123} + \, |111\rangle_{123})
\ee
  is suitable for perfect teleportation and superdense coding, there
  are states in its category which are not suitable without the
  application of SLOCC. Similarly, although prototype W-state
\be
|W \rag_{123} = \sqrtthr \,(|100\rag_{123} + \, |010\rangle_{123} + 
                      \, |001\rangle_{123})
\ee
  may not be suitable for perfect teleportation and superdense 
  coding, but there may exist class of states within W-states
  category which is suitable. Below, we give one such example.

  The class of states $|W_{n} \rag_{123}$ belongs to the category 
  of W-states that can be used as an entanglement resource. This is
given by 
\be
|W_{n} \rag_{123} = \norm \,(|100\rag_{123} + 
                    \, \sqrt{n} e^{i\gamma} |010\rangle_{123} +
		   \, \sqrt{n+1} e^{i\delta}|001\rangle_{123}),
\ee
 where $n$ is a real number and $\gamma$ and $\delta$ are phases.
 As we shall see this class of W-states can be used for 
 perfect teleportation and superdense coding.
In particular, if we take $n=1$ for simplicity and set phases to
zero, then we have the following W-state
\be
|W_{1} \rag_{123} = \half \,(|100\rag_{123} + \, |010\rangle_{123} +
                          \, \sqrt{2}|001\rangle_{123}).
\ee
We can check that this state belongs to the category of 
W-states using the criteria given in the Ref \cite{dvc}. This state 
has true tripartite entanglement. This is because the 
determinants of the reduced density matrices $\rho_{1}, 
\rho_{2}$ and $\rho_{3}$ of this state are not zero. 
For a state belonging to GHZ category 3-tangle
is non-zero; while it is zero for a state belonging to
the W-state category. For the state $|W_{1} \rag_{123}$
(and $|W_{n} \rag_{123}$),
the 3-tangle is zero. (The prescription to compute 3-tangle
is given in Ref. \cite{ckw}). We also note that the concurrences 
$C_{12} = \half$, $C_{13} = \sqrttwo$ and 
$C_{1(23)} = {\sqrt{3} \over 2}$. This is because
the concurrence $C_{ab} = 2 \sqrt{Det \rho_a}$ for a 
pure state of the system `ab'. So the inequality of the
Ref. \cite{ckw}, $C_{12}^{2} + C_{13}^{2} \leq C_{1(23)}^{2}$ is 
saturated and there is no residual entanglement. This is 
a characteristic of a state belonging to the W-state
category.

  Another way to argue that W-state is not
  equivalent to a GHZ state is to compute reduced
  density matrices $\rho_{12}, \rho_{23}$ or $\rho_{13}$.
  We can check that the reduced density matrices obtained from 
$|W_{1} \rag_{123}$ cannot be written
  as a linear combination of 
  product states, while for the $|GHZ \rag_{123}$ they can be. For example, 
for the $|GHZ \rag_{123}$ state we have 
\be
   \rho_{12} = \half (|00 \rag \lag 00| + |11 \rag \lag 11|),
\ee
   while for the $|W_{1} \rag_{123}$ state we have 
\be
\rho_{12} = {1 \over 3} (|00 \rag \lag 00| + 2|\psi^{+} \rag \lag\psi^{+}|),
\ee
where $|\psi^{+} \rag = \half (|00 \rag + |11 \rag)$ is a Bell state.
Two such states cannot be equivalent under SLOCC.

We can use either the state given in (3) or (4) for the purpose of
perfect quantum teleportation. In our scheme Alice will have two qubits and
Bob will have one qubit. Then, by performing suitable projection on the 
input qubit and the two qubits, Alice can convey the measurement outcome 
to Bob, who can then perform suitable local unitary transformation to convert
his particle into an unknown state. One may ask how much entanglement is shared
between Alice and Bob? Even though, the shared W-state is a three qubit state,
but with respect to Alice and Bob partitioning we can imagine this 
as a bipartite system. Then, we can calculate the von Neumann entropy of the 
either reduced subsystem to know how much entanglement is there between 
the subsystem of Alice and the subsystem of Bob \cite{pr}. Now, 
if we trace the particle `1' and `2' then the reduced density 
matrix of particle `3' is random mixture for all values of $n$, i.e., 
$\rho_3 = {\rm tr}_{12} (|W_n \ra\la W_n|) = I/2$. This shows that the von 
Neumann entropy of $\rho_3$ is just one. That is there is one ebit of 
entanglement shared between Alice and Bob when they use this class of 
W-state. As we will see, by using one ebit of shared entanglement and 
communication of two classical bits Alice can send a qubit to Bob using
W-state as a resource.

\vspace{0.05in}

\section{Teleportation with W-states}

    Let us consider a situation where Alice has particles
    `1' and `2' and Bob has the particle `3'. Alice wishes
    to teleport the unknown state of a particle `a'. She
    can make a measurement on the three particles `a12'
    and convey her results to Bob via classical communication.
    Then the question is: what are the states that Alice and
    Bob can share that can be used as a quantum resource to
    transfer the state of particle `a' to particle `3' using
    usual teleportation protocol. 
    
      Before discussing the protocol using a W-state as a resource, 
    let us describe how does the protocol works for the prototype 
    GHZ state. Suppose that Alice and Bob share a three qubit
  entangled state $|GHZ \rag_{123}$ given by 
\be
|GHZ \rag_{123} = \sqrttwo \,(|000\rag_{123} + \, |111\rangle_{123}).
\ee   
Alice has particles
  `1' and `2', while the Bob has the particle `3'.
  Alice also has a particle `a' in the following unknown state

\be
|\psi \rag_{a} = (\alpha|0\rag_{a} + \, \beta |1\rangle_{a}).
\ee
Alice now wishes to transmit this state to Bob. To see how
the protocol works, let us consider the combined input and entangled 
state and rewrite it as
\bea
|\psi\rag_{a} |GHZ \rag_{123} & = &  \sqrttwo (\alpha |0\rag_{a} + \, \beta |1\rag_{a})\,
(|000\rag_{123} + \, |111\rangle_{123}) \nonumber  \\
  & = & \half  [\,|\psi^{+}_{1}\rag_{a12} \, (\alpha |0\rag_{3}
          + \beta |1\rag_{3} ) + \;|\psi^{-}_{1} \rag_{a12}
         (\alpha |0\rag_{3} - \beta |1\rag_{3} ) \nonumber  \\
	&    & \; + \; |\psi^{+}_{2}\rag_{a12}
         (\beta |0\rag_{3} + \,\alpha |1\rag_{3} ) 
	+  \; |\psi^{-}_{2} \rag_{a12} 
	 (\,\beta|0\rag_{3} - \,\alpha |1\rag_{3} ) ].
\eea
Alice can make a von-Neumann type measurement using the states 
   $\{|\psi^{\pm}_{1}\rag, |\psi^{\pm}_{2}\rag \}$, which are given by 
\bea
    |\psi^{\pm}_{1}\rag & = & \sqrttwo \,(|000\rag \pm \,|111\rag),  
	                                            \nonumber  \\
    |\psi^{\pm}_{2}\rag & = & \sqrttwo \,(|100\rag \pm \,|011\rag).  
\eea
After performing three particle measurement, 
   Alice can convey her results to Bob by sending two classical bits
   of information. Bob can then convert the state of his particle
   to that of particle `a' by applying appropriate unitary
   transformations.

   Let us examine the Category of W-states.
 The prototype example of this category is given by 
   
\be
|W \rag_{123} = \sqrtthr \,(|100\rag_{123} + \, |010\rangle_{123} + 
                      \, |001\rangle_{123}).
\ee
This state is not useful as a quantum resource for the usual
teleportation protocol \cite{gor}.
But it turns out that we can use a class of states of the W-state
   category for teleportation. An example of this class of states
   is given by 

\be
|W_{1} \rag_{123} = \half \,(|100\rag_{123} + \, |010\rangle_{123} +
                          \, \sqrt{2}|001\rangle_{123}).
\ee
Let us suppose that Alice and Bob share a three qubit
  entangled state $|W_{1} \rag_{123}$. Alice has particles
  `1' and `2', while the Bob has the particle `3'.
  Alice also has a particle `a' in the unknown state $(8)$.
Alice now wishes to teleport this state to Bob. To examine its
possibility, let us consider the combined input and entangled 
state and rewrite this as

\bea 
|\psi\rag_{a} |W_{1} \rag_{123} & = &  \half (\alpha |0\rag_{a} + \, \beta |1\rag_{a})\,
(|100\rag_{123} + \, |010\rangle_{123} + \, \sqrt{2}|001\rangle_{123})
  \nonumber  \\
  & = & \half \; [\alpha |010\rag_{a12} 0\rag_{3} + \alpha |001\rag_{a12}  |0\rag_{3} +
     \, \sqrt{2} \alpha \, |000\rag_{a12}|1\rag_{3} \nonumber \\ 
      &    &  \; + \;
    \; \beta |110\rag_{a12} |0\rag_{3} + \beta |101\rag_{a12} |0\rag_{3} +
     \, \sqrt{2} \beta \, |100\rag_{a12}|1\rag_{3} ] 
  \nonumber  \\
  & = & \half \; [\alpha (|010\rag_{a12} + |001\rag_{a12} ) |0\rag_{3} +
     \, \sqrt{2} \alpha \, |000\rag_{a12}|1\rag_{3} \nonumber \\
  &    &  \; + \;
     \; \beta (|110\rag_{a12} + |101\rag_{a12} ) |0\rag_{3} +
     \, \sqrt{2} \beta \, |100\rag_{a12}|1\rag_{3} ] 
  \nonumber  \\
  & = & \half  [\,|\eta^{+}_{1}\rag_{a12} \, (\alpha |0\rag_{3}
          + \beta |1\rag_{3} ) + \;|\eta^{-}_{1} \rag_{a12}
         (\alpha |0\rag_{3} - \beta |1\rag_{3} ) \nonumber  \\
	&    & \; + \; |\xi^{+}_{1}\rag_{a12}
         (\beta |0\rag_{3} + \,\alpha |1\rag_{3} ) 
	+  \; |\xi^{-}_{1} \rag_{a12} 
	 ( \,\beta|0\rag_{3} - \,\alpha |1\rag_{3} ) ] \nonumber \\
&=& \half [|\eta^{+}_{1}\rag_{a12} \otimes |\psi\rag_3 + 
|\eta^{-}_{1} \rag_{a12} \otimes \sigma_3 |\psi\rag_3 + 
|\xi^{+}_{1}\rag_{a12} \otimes \sigma_1 |\psi\rag_3  \nonumber \\
 &  &  + |\xi^{-}_{1} \rag_{a12} \otimes (-i\sigma_2)|\psi\rag_3 ]. 
\eea
Here $|\eta^{\pm}_{1}\rag$ and $|\xi^{\pm}_{1}\rag$ are a set 
   of orthogonal states in the W-state category given by 
\bea
        |\eta^{\pm}_{1}\rag & = & \half \,(|010\rag + \,|001\rag \pm 
	                \, \sqrt{2} |100\rag),   \\
        |\xi^{\pm}_{1}\rag & = & \half \,(|110\rag + \,|101\rag \pm 
	                \, \sqrt{2} |000\rag). 
\eea
Alice can now make a von Neumann measurement in a basis
  that includes the states
  $\{|\eta^{\pm}_{1}\rag, |\xi^{\pm}_{1}\rag \}$ on the
  combined system of three particles `a12'. She then sends
  the result of her measurements using two classical 
  bits to Bob who can apply one of the unitary 
  transformations  $\{ 1, \sigma_{1}, 
  i\sigma_2, \sigma_3 \}$ to convert the state of his particle 
  `3' to that of the particle `a'. For example, if the outcome is
$|\xi^{+}_{1} \rag_{a12}$, then Bob has to apply  $\sigma_1$ to get 
the desired state $|\psi\rag_3$. This completes the teleportation protocol 
using W-state. This protocol consumes one ebit of shared entanglement and 
communication of two classical bits of information between Alice and Bob.
    
As mentioned before, the state $|W_{1} \rag_{123}$ belongs to the 
following class of W-states that can also be used as an entanglement resource

\be
|W_{n} \rag_{123} = \norm \,(|100\rag_{123} + 
                    \, \sqrt{n} e^{i\gamma} |010\rangle_{123} +
		   \, \sqrt{n+1} e^{i\delta}|001\rangle_{123}).
\ee
Alice then can use a set of basis vectors for the measurement
   that includes the vectors:
\bea
|\eta^{\pm}_{n}\rag & = & \norm \,(|010\rag + 
                          \sqrt{n} e^{i\gamma}\,|001\rag \pm 
			  \, \sqrt{n+1} e^{i\delta}|100\rag),   \\
    |\xi^{\pm}_{n}\rag & = & \norm \,(|110\rag +
                              \sqrt{n} e^{i\gamma}  \,|101\rag \pm 
			      \,\sqrt{n+1} e^{i\delta}|000\rag) .
\eea
Using these orthogonal states Alice can carry out the quantum 
teleportation protocol as given above.
  
As a remark, the GHZ state, given in $(1)$,
  can not only be used for the above situation, but for other
  scenarios also. We can envisage another situation.
  Alice instead of making a three-particle measurement, may wish
  to make a two particle measurement, followed by an one-particle
  measurement. Or, there can be three parties: Alice, Bob and Charlie,
  each with a qubit. Then Alice makes a two particle measurement
  on `a1' and Bob makes a one particle measurement on the
  particle `2' and the state of the particle `a' is teleported 
  to Charlie who makes appropriate unitary
  transformations on his qubit to change the state of his particle
  `3' to that of the particle `a'. The above GHZ state can be used
  in this scenario as a resource \cite{kb,akp}; 
but the above $|W_{n} \rag_{123}$
  state may not work for this scenario in a straightforward
  manner.

\vspace{0.05in}

\section{Superdense Coding and W-states}

     In the original superdense coding scenario, Alice can transmit
     two classical bits to Bob by sending a qubit if they share a Bell 
state \cite{bw}. First, Alice encodes her message by making suitable unitary
     transformation on her qubit and then sends back it to Bob. Bob has 
two qubits at his disposal who can perform von Neumann projections to 
distinguish four Bell-states and retrive the two classical bits.

     In this section we briefly discuss how a GHZ state can be used 
for superdense coding. If Alice
     and Bob share such a state then Alice can transmit two
     classical bits by sending one qubit. We can see this 
     as follows. The prototype GHZ state (1) is same as
     the state  $|\psi^{+}_{1}\rag$. Let Alice has particle `1'
     and Bob has the particles `2' and `3'. Alice can apply
     $\{I, \sigma_{1}, i \sigma_{2}, \sigma_{3}\}$ transformations
     on her qubit

\bea
|\psi^{+}_{1}\rag \rightarrow I \otimes I \otimes I |\psi^{+}_{1}\rag 
= |\psi^{+}_{1}\rag \nonumber\\
|\psi^{+}_{1}\rag \rightarrow  \sigma_1 \otimes I \otimes I 
|\psi^{+}_{1}\rag = |\psi^{+}_{2}\rag \nonumber\\
|\psi^{+}_{1}\rag \rightarrow i\sigma_2 \otimes I \otimes I 
|\psi^{+}_{1}\rag = |\psi^{-}_{2}\rag \nonumber\\
|\psi^{+}_{1}\rag \rightarrow \sigma_3 \otimes I \otimes I 
|\psi^{+}_{1}\rag = |\psi^{-}_{1}\rag.
\eea
These transformations convert the original state to a set of four
orthogonal states: $\{|\psi^{\pm}_{1}\rag, |\psi^{\pm}_{2}\rag \}$. After
receiving Alice's qubit, Bob can make three-particle measurement using
these orthogonal states to recover two classical bits.
 However, using $|W \rag_{123}$ state Alice cannot transmit two 
     classical bits by sending one qubit by using
     conventional superdense coding protocol.

     The advantage of the state $|W_{n} \rag_{123}$ is that
     using a scheme analogous to the usual superdense coding protocol, 
Alice can transmit
     two classical bits to Bob by sending one qubit. For this state,
     the protocol works as follows: Alice has the qubit `1',
     while Bob has the qubits `2' and `3'. For simplicity,
     let Alice and Bob share $|\eta_{+} \rag_{123}$, which is
     a $|W_{n} \rag_{123}$ type state. Alice
     can now apply $\{I, \sigma_{1}, i \sigma_{2}, \sigma_{3}\}$
     on her qubit `1'. This will lead to the set of four orthogonal
     states as given below
\bea
|\eta^{+}_{1}\rag \rightarrow I \otimes I \otimes I |\eta^{+}_{1}\rag 
= |\eta^{+}_{1}\rag \nonumber\\
|\eta^{+}_{1}\rag \rightarrow \sigma_1 \otimes I \otimes I 
|\eta^{+}_{1}\rag = |\xi^{+}_{1}\rag \nonumber\\
|\eta^{+}_{1}\rag \rightarrow i\sigma_2 \otimes I \otimes I 
|\eta^{+}_{1}\rag = |\xi^{-}_{1}\rag \nonumber\\
|\eta^{+}_{1}\rag \rightarrow \sigma_3 \otimes I \otimes I 
|\eta^{+}_{1}\rag = |\eta^{-}_{1}\rag,
\eea
where the set of four mutually orthogonal basis are 
$\{|\eta^{\pm}_{1}\rag, |\xi^{\pm}_{1}\rag \}$. These are defined
in $(14)$ and $(15)$.

  Now Alice can send her qubit to Bob who makes a three-particle
    von-Neumann measurement using only the orthogonal states
    $\{|\eta^{\pm}_{1}\rag, |\xi^{\pm}_{1}\rag \}$. Since these are 
orthogonal, Bob can perfectly distinguish what operation Alice has applied. 
In this way
    he can recover two classical bit of information. Here, it is 
    necessary that Bob knows the shared state in advance; then
    he will know about the orthogonal states to use for the measurement.
    More generally, if Alice and Bob have shared the entangled state 
$|W_n\ra$ given in (3), then they can use the orthogonal W-states
    $\{|\eta^{\pm}_{n}\rag, |\xi^{\pm}_{n}\rag \}$ for superdense coding.
The superdense coding protocol with W-state also uses one ebit of shared 
entanglement.

      The GHZ state $(1)$ can also be used to transmit three classical
    bits by sending two qubits. This works as follows. Alice has now
    particles `1' and `2', while Bob has the particle `3'. Alice
    can make unitary transformation on both her particles. Then apart
    from the transformations given in $(19)$, she can apply the following 
transformations

\bea
|\psi^{+}_{1}\rag \rightarrow I \otimes \sigma_1  \otimes I 
|\psi^{+}_{1}\rag = |\psi^{+}_{3}\rag \nonumber\\
|\psi^{+}_{1}\rag \rightarrow I \otimes i \sigma_2  \otimes I 
|\psi^{+}_{1}\rag = |\psi^{-}_{3}\rag \nonumber\\
|\psi^{+}_{1}\rag \rightarrow \sigma_1 \otimes \sigma_1  \otimes I 
|\psi^{+}_{1}\rag = |\psi^{+}_{4}\rag \nonumber\\
|\psi^{+}_{1}\rag \rightarrow \sigma_1 \otimes i \sigma_2  \otimes I 
|\psi^{+}_{1}\rag = |\psi^{-}_{4}\rag, 
\eea
 where $\{|\psi^{\pm}_{3}\rag, |\psi^{\pm}_{4}\rag \}$ are given by 
\bea
    |\psi^{\pm}_{3}\rag & = & \sqrttwo \,(|010\rag \pm \,|101\rag),  
	                                            \nonumber  \\
    |\psi^{\pm}_{4}\rag & = & \sqrttwo \,(|110\rag \pm \,|001\rag).  
\eea

Now Alice can send her two qubits to Bob, who makes von Neumann
   measurement using the orthogonal set 
$\{|\psi^{\pm}_{1}\rag, |\psi^{\pm}_{2}\rag, |\psi^{\pm}_{3}\rag, 
|\psi^{\pm}_{4}\rag \}$. This measurement will yield three classical bit of 
information. 
Unfortunately, it may not be possible to use the W-states 
$|W_{n} \rag_{123}$ to
transmit three classical bits of information. But, it is possible that
some other class of W-states exist which can be used for such a task.

    As earlier, we would note that $|W_{n} \rag_{123}$ may not 
    be only class of W-states that allow superdense coding. If
    one explores further, specifically those W-states that are
    linear superposition of more than three states, one may find
    W-states which may even allow Alice to transmit three classical
    bits by sending two qubits. This is worth exploring in future. 

\vspace{0.05in}

\section{Conclusions}

In this paper we have shown that there exists a class of W-states which are 
useful for quantum teleportation. The scheme presented here works analogously
to the original protocol. The only difference is that Alice needs to carry 
out a 
three-qubit von Neumann projection instead of a Bell-state measurement. 
Interestingly, the resource used for teleporting an unknown qubit using 
the W-state
is one shared ebit and communication of two classical bits. This reassures
that we cannot outperform quantum teleportation of a qubit 
with less resource. The 
original protocol also uses one shared ebit and two classical bits.
Furthermore, we have shown that one can perform standard superdense coding
scheme using a W-state. Again, here by sharing one ebit and sending a qubit 
Alice can communicate two classical bits to Bob. So instead of using EPR
pair for quantum teleportation and dense coding one can use ${\rm W}_n$-state 
as well for the above purpose. In future, one can investigate if there 
are other classes of W-states useful for quantum teleportation and 
superdense coding. One can 
also study if there are classes of W-states that can be used to send 
three classical bits by sending two qubits. One can also investigate
the possibility of teleporting an unknown state by making two-qubit
and one-qubit measurements instead of a three-qubit measurement with 
a class of W-states as a resource.

\end{document}